# Nondestructive 3D Metrology of Through-Glass Via Geometry and Internal Wall Texture Using Holotomography

**Dohyeon Lee[1,2], Herve Hugonnet[1,2] and YongKeun Park[1,2,3]**

[1]Department of Physics, Korea Advanced Institute of Science and Technology (KAIST), Daejeon 34141, Republic of Korea
[2]Institute of Imaging, KAIST, Daejeon 34141, Republic of Korea
[3]Tomocube, Daejeon 34127, Republic of Korea
Corresponding author: yk.park@kaist.ac.kr

ABSTRACT Through-glass vias (TGVs) are increasingly important for semiconductor packaging, yet their internal geometry and wall texture remain difficult to characterize nondestructively. Here, we use holotomography (HT) to inspect the three-dimensional (3D) geometry and internal wall texture of intact, unfilled TGVs before metallization. HT reconstructs the refractive index (RI) distribution of the glass and via. Maps of RI and spatially filtered RI then localize the internal boundary. Geometric fitting quantifies via dimensions and taper, while spatial filtering separates slowly varying wall form from finer texture. The reconstructed surface supports volumetric visualization, dimensional measurements and spatial characterization of wall texture within one optical inspection workflow. These measurements reveal internal features inaccessible from surface views and provide quantitative descriptors to evaluate TGVs.

## I. INTRODUCTION

Through-glass vias (TGVs) provide vertical interconnections in glass substrates for semiconductor packaging [1], [2] (Fig. 1a). Laser modification and selective etching form through-holes and high-aspect-ratio structures [3], [4], [5], [6], [7], [8], [9]. Lithography-free and ultrashort-pulse approaches extend these fabrication routes [10], [11]. Characterizing the internal geometry and wall texture before copper filling is a key inspection task. Studies of copper/glass adhesion, roughness-dependent microwave performance and cracking in copper-filled vias motivate inspection at this stage [12], [13], [14], [15]. Surface views reveal the openings, whereas cross-sectional inspection requires destructive sample preparation. The challenge is therefore to access the internal waist and sidewall while preserving the intact via.

Optical methods measure internal via dimensions and wall profiles through several imaging approaches. Holographic TGV metrology addresses morphology and wall profiles using two-dimensional optical-field measurements [16]; shadowgraphy resolves top, waist and bottom geometry [17]. Optical coherence tomography measures microhole depth and diameter [18], and spectral-coherence methods measure transparent blind-glass-via depth [19]. These studies motivate a volumetric measurement that resolves overall via geometry and finer wall features across the internal surface.

Here, we use holotomography (HT) for the inspection of TGVs. HT reconstructs three-dimensional (3D) refractive index (RI) distributions [20] from multiple two-dimensional measurements [21], [22], [23], [24], [25], [26], [27]. Intensity diffraction tomography, multiplexed illumination, transport-of-intensity methods and low-coherence angle scanning offer different routes to RI reconstruction [28], [29], [30], [31], [32]. HT has also supported imaging of lenses, indented glass and volume holographic gratings [33], [34], [35]. Regularization further improves reconstruction under incomplete optical support [36], [37], [38].

Here, we demonstrate nondestructive HT measurement of the full 3D geometry and internal wall texture of intact, unfilled TGVs. Boundary extraction, geometric fitting and spatial filtering quantify top, waist and bottom diameters, taper angles, axial length and sidewall roughness. We apply this approach to 192 vias from two substrate classes and evaluate sidewall-roughness consistency using 188 physically inverted pairs and 134 HT–micro-computed tomography (micro-CT) pairs. The resulting surface links volumetric visualization with measurements of via form and wall texture before metallization.

## II. RESULTS

### *A. Volumetric inspection of the internal via surface*

Four light-emitting diode (LED) illumination measurements are acquired at each axial position. Deconvolution with a 3D point-spread function (PSF) converts these intensity measurements into a 3D RI image [22], [23], [24] (Fig. 1b). Total variation (TV) refinement reduces noise and constrains the reconstruction under incomplete angular support. We used an $l_2$ norm of the discrete gradient and a TV weight of $10^{-3}$ (Supplementary Fig. S2). Extracting the boundary enables measurements of diameter, taper and length, spatial wall-texture maps and sidewall roughness (Fig. 1c).

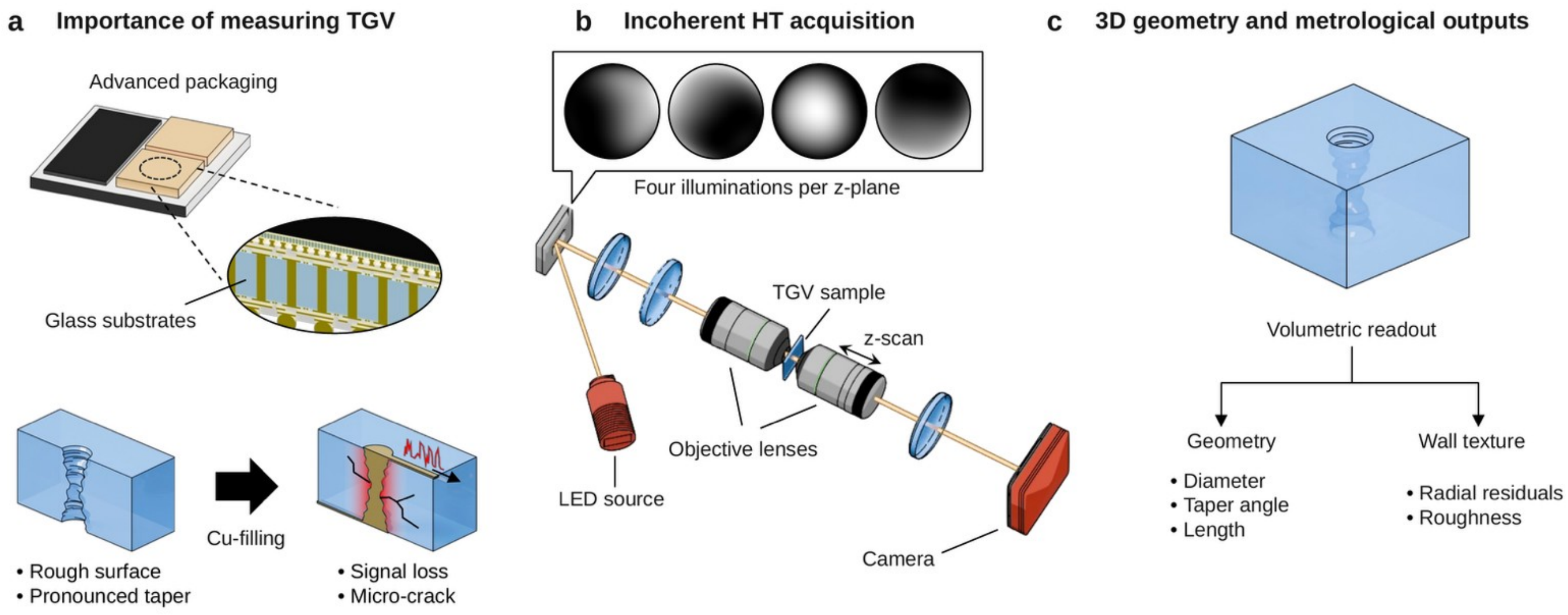

**FIGURE 1.** Schematics of nondestructive inspection of internal through-glass via (TGV) geometry and wall texture. (a) Packaging context and the motivation for inspecting unfilled TGVs before copper filling. Surface irregularity and tapered shape are potential concerns in a filled interconnect. (b) Holotomography (HT) acquisition using light-emitting diode (LED) illumination, four intensity measurements per axial plane and axial stage scanning. (c) Three-dimensional (3D) structure and the geometric and wall-texture measurements obtained from its boundary. Cu denotes copper, and z denotes the axial coordinate.

### *B. Refractive index matching and boundary extraction*

High RI contrast between the sample and background medium can cause strong refraction and distort the reconstructed image. RI matching changes the immersion medium to reduce this contrast [33]. In TGVs, matching oils reduced visible distortion at the via wall while retaining contrast for boundary extraction (Fig. 2b). After testing the immersion series (Supplementary Fig. S1), the selected oil indices were 1.500 for the low-roughness (LR) substrate and 1.465 for the high-roughness (HR) substrate. Supplementary Fig. S1 also shows toluene immersion as an alternative tested condition.

Segmentation of the RI volume identifies the lateral via boundary, but axial localization is less direct. Morphology-preserving holotomography (MP-HT) applies spatial-frequency filtering to the RI reconstruction to emphasize morphology [39].

Here, MP-HT assists localization of the axial openings (Fig. 2a,c). In the synthetic example, full-volume 3D mask intersection over union (IoU) increased from 0.93 for RI-based Otsu segmentation to 0.96 with MP-HT-assisted endpoint localization. In experimental LR and HR examples, endpoint localization removed segmented regions extending into the immersion liquid (Supplementary Fig. S3). The synthetic IoU measures volume overlap rather than boundary-position accuracy.

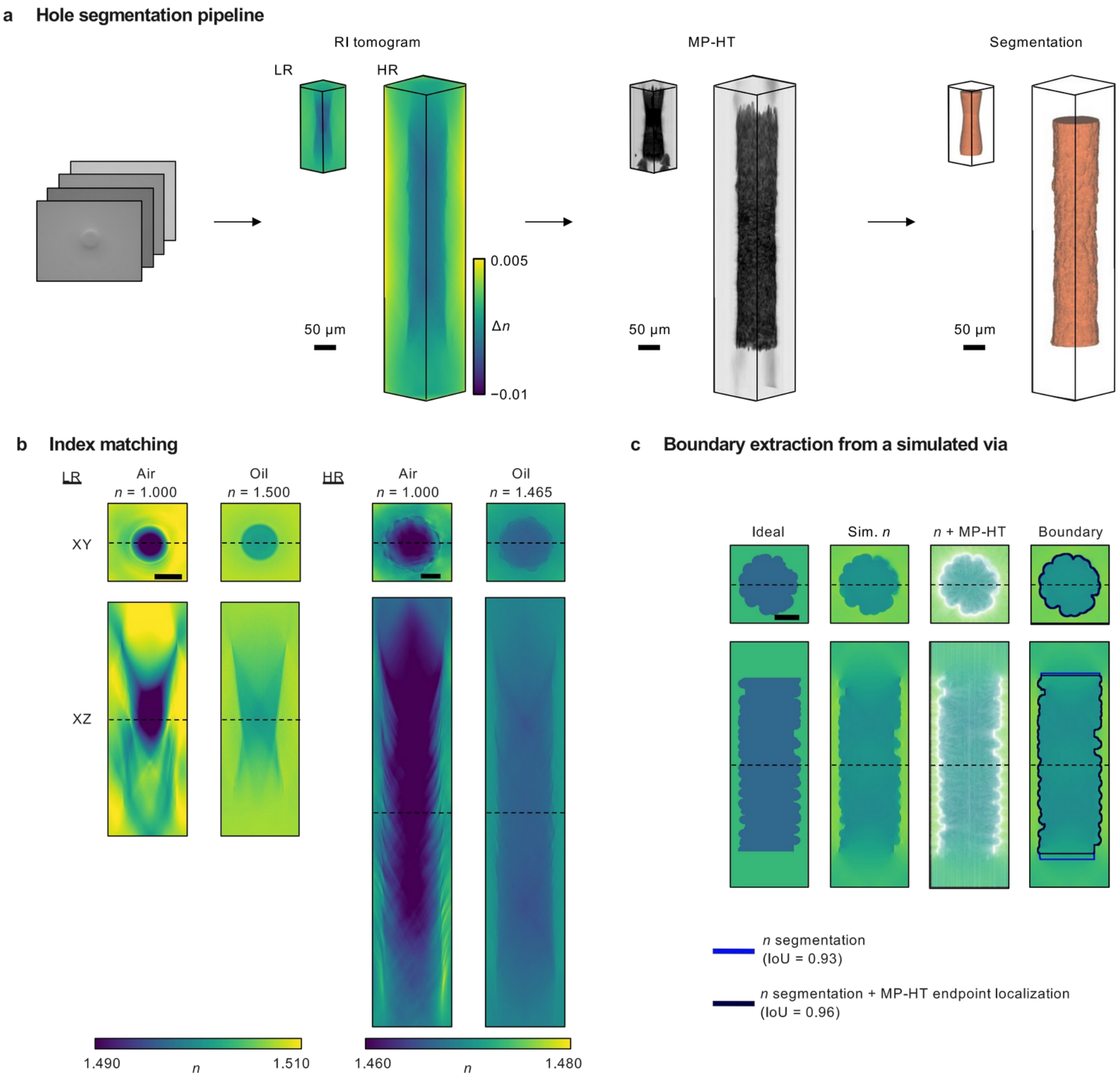

**FIGURE 2.** Refractive index (RI) matching and extraction of the internal via boundary using holotomography (HT). (a) Reconstruction and segmentation pipeline for representative smaller low-roughness (LR) and larger high-roughness (HR) vias. The RI volume is transformed into a morphology-preserving holotomography (MP-HT) representation, which assists endpoint localization during segmentation. Renderings within each class share the region, orientation and display scale; scale bars, 50 µm. The three-dimensional (3D) RI maps are displayed as $\Delta n = n - n_{bg}$, where $n$ denotes RI and $n_{bg}$ is the reconstruction-background index (1.470 for HR and 1.507 for LR). (b) Transverse and longitudinal sections in air or the selected oils ($n$ = 1.500 for LR and 1.465 for HR). (c) Synthetic reference, simulated RI reconstruction, RI combined with MP-HT, and extracted boundaries. The full-volume 3D mask intersection over union (IoU) is 0.93 for RI-based segmentation and 0.96 with MP-HT-assisted endpoint localization. Scale bars in b and c, 25 µm. Supplementary Figs. S1–S3 show the extended comparisons. IoU is a volume-overlap statistic and does not quantify boundary-position error. The label Sim. denotes simulated reconstruction.

### *C. Separation of via form and wall texture*

The macroscopic via geometry can be described by a straight centerline and a piecewise-linear hourglass profile (Fig. 3a). Local circle fits give the top, waist and bottom diameters, while the hourglass fit gives the waist position and upper and lower taper. Subtracting the hourglass reference radius from the measured wall radius gives a residual height indexed by angle and depth. The residual height contains fine wall texture information together with gradual radius changes that are not captured by the reference.

A Gaussian L-filter separates the residual into a slowly varying waviness surface and a filtered-height map (Fig. 3a,b). Sidewall roughness is the mean absolute filtered height (Methods E).

Two examples in Fig. 3b illustrate this separation. Filtering reveals pit-like texture in the HR example, whereas the LR filtered height is nearly uniform at the displayed color scale. In the HR example, the top feature and gradual axial trend are largely assigned to waviness, while the filtered map retains fine pit-like texture distributed over the wall. The displayed values describe individual examples; population summaries follow below.

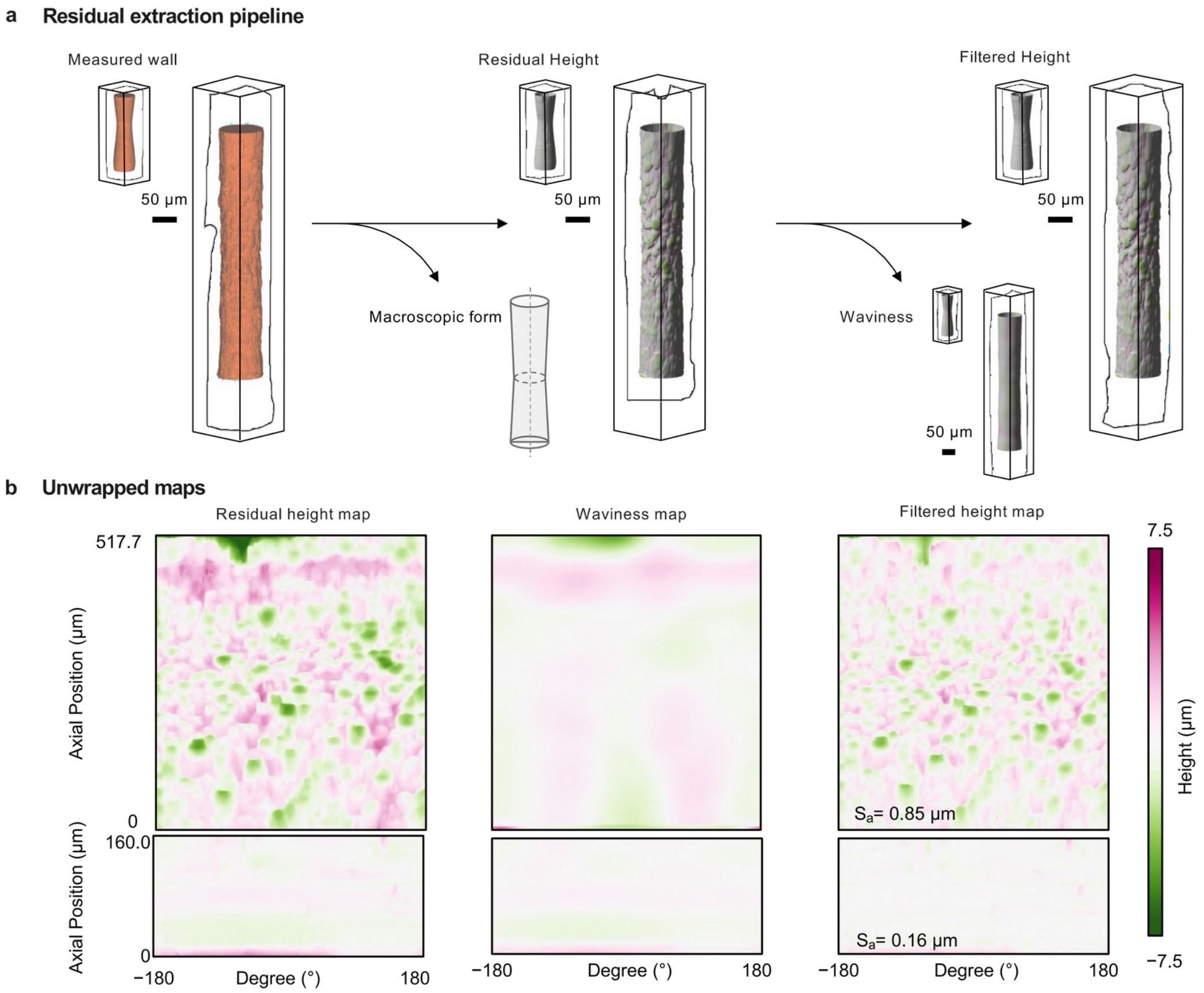

**FIGURE 3.** Separation of macroscopic via form, waviness and wall texture. (a) Surface-processing workflow for representative smaller low-roughness (LR) and larger high-roughness (HR) vias. Subtraction of the fitted hourglass form gives the residual height; removal of Gaussian-filtered waviness gives the filtered height. Scale bars, 50 μm. (b) Unwrapped residual-height, waviness and filtered-height maps for HR (top) and LR (bottom). Axial lengths are 517.7 and 160.0 μm, and representative sidewall roughness values are 0.85 and 0.16 μm, respectively. All maps share the same signed-height color scale. Sidewall roughness is denoted by $S_a$.

### *D. Population measurements of via geometry and roughness*

The population of vias used for geometrical analysis comprised 192 vias: 100 HR and 92 LR. The HR vias had larger top, waist and bottom diameters than LR vias, consistent with their larger nominal dimensions (Fig. 4). Taper angles closer to 90° indicate more vertical HR walls; LR vias showed a more pronounced hourglass profile. Figure 4a,b summarizes these geometric distributions.

Mean segmented axial lengths were close to the bulk caliper reading for HR, whereas the LR mean was higher (Supplementary Fig. S5 and Table S1). These bulk readings were not co-located with individual via measurements and do not establish traceable axial accuracy.

The LR vias had lower mean sidewall roughness than the HR vias, with greater dispersion relative to their mean (Fig. 4c). The mean values were 0.272 μm for LR and 0.841 μm for HR; sample standard deviations (SDs) are listed in Supplementary Table S1.

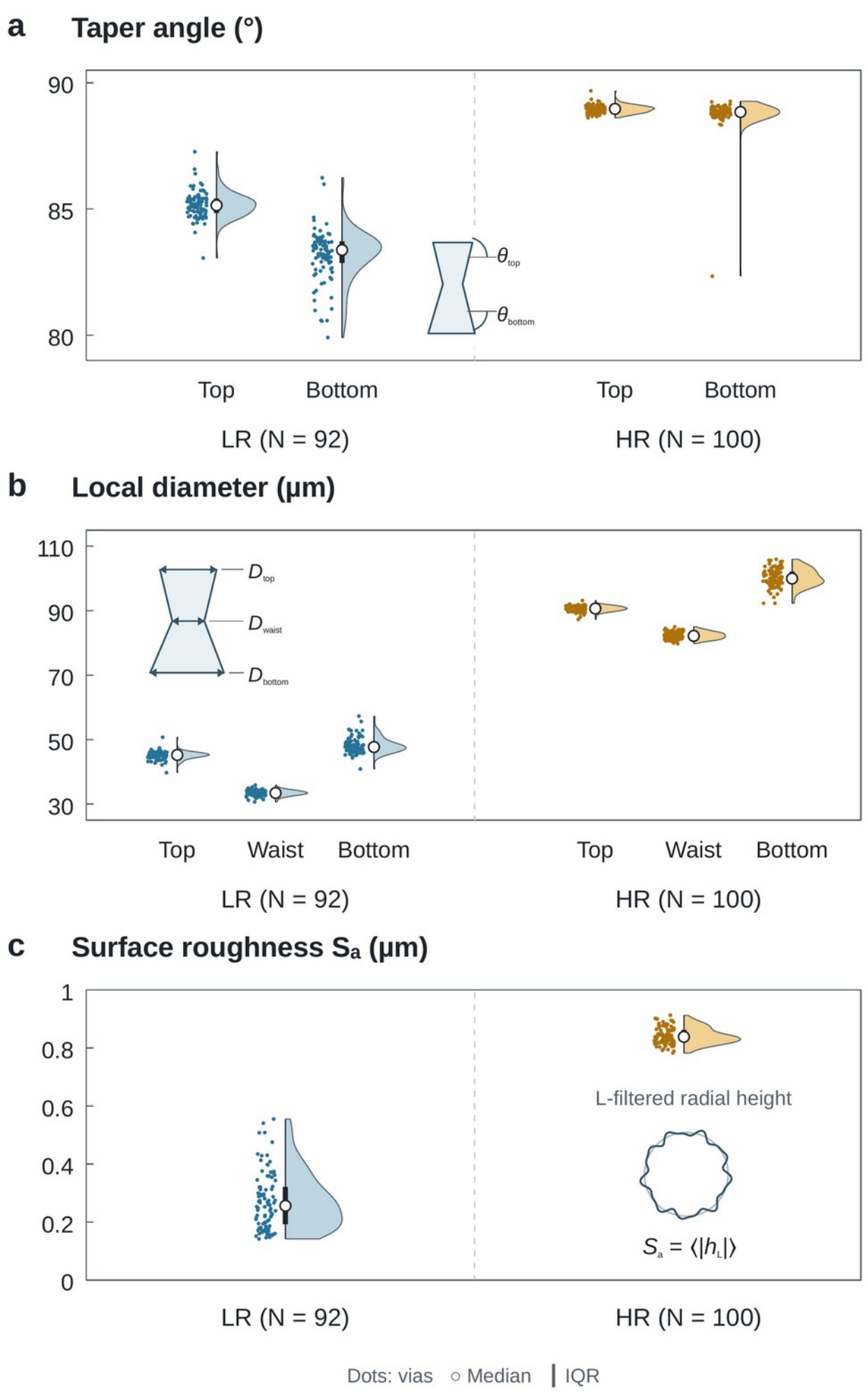

**FIGURE 4.** Population measurements of via geometry and sidewall roughness for 92 low-roughness (LR) and 100 high-roughness (HR) vias in the reference orientation, with schematic insets defining the measured quantities. (a) Top and bottom taper angles relative to the transverse plane, $\theta_{top}$ and $\theta_{bottom}$. (b) Local top, waist and bottom diameters, $D_{top}$, $D_{waist}$ and $D_{bottom}$. The waist is the slice selected by the hourglass fit. (c) Sidewall roughness, $S_a$, calculated as the mean absolute L-filtered height. Each colored point represents one via; horizontal jitter is used only for visibility. Right half-violins show kernel density estimates, truncated to the observed range and normalized to equal maximum width within each panel. White circles indicate medians, thick black segments span the interquartile range, and thin black lines extend from the observed minimum to maximum. Blue and ochre denote LR and HR, respectively. Insets are conceptual and are not to scale; the circular contour is illustrative. N denotes the number of vias.

### *E. Orientation and micro-computed tomography comparisons*

To assess sensitivity to orientation and remounting, we compared reference HT measurements with measurements after physical inversion. A separate comparison with micro-CT assessed modality-dependent differences.

Roughness showed little average shift after physical inversion and remounting (Fig. 5a,b). Across 188 matched vias, the mean roughness difference of the reference HT versus flipped HT was +0.002 µm. The limits of agreement (LoA) quantify the variation between individual measurements in this orientation and remounting test (Supplementary Table S2).

The comparison with micro-CT comprised 134 matched vias. Roughness difference in reference HT relative to micro-CT depended on substrate class (Fig. 5a,c): micro-CT gave lower values for HR and higher values for LR. The class-specific mean differences were +0.124 and −0.164 µm, respectively. Supplementary Table S2 provides the paired means and agreement statistics; Supplementary Fig. S4 shows transverse and longitudinal views.

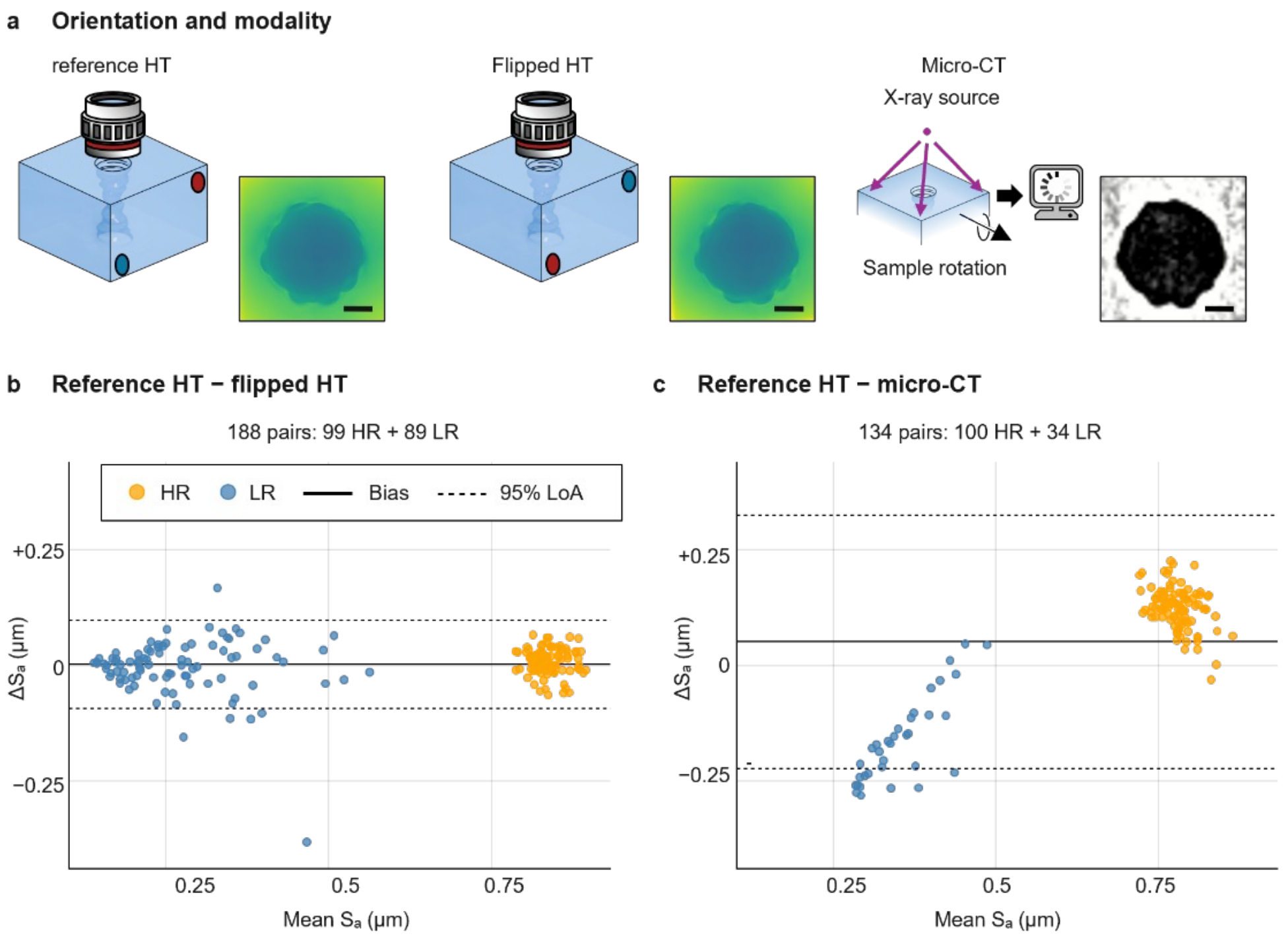

**FIGURE 5.** Orientation and micro-computed tomography (micro-CT) comparisons. (a) Reference and physically flipped holotomography (HT) measurements, and micro-CT with sample rotation, with representative high-roughness (HR) transverse images. Scale bars, 25 µm. (b) Bland–Altman comparison of reference HT and flipped HT measurements of sidewall roughness $S_a$ for 188 pairs (99 HR and 89 low-roughness (LR) vias). The pooled bias is +0.002 µm and the 95% limits of agreement (LoA) are −0.094 and +0.097 µm; six points lie outside them. (c) Bland–Altman comparison of reference HT and micro-CT for 134 pairs (100 HR and 34 LR), with bias +0.051 µm and LoA −0.223 and +0.325 µm; 11 points lie outside. Class-specific mean differences are +0.124 µm for HR and −0.164 µm for LR. The horizontal coordinate is the pair mean; the vertical coordinate $\Delta S_a$ is reference HT minus the comparison measurement. Solid and dashed lines indicate pooled bias and bias ±1.96 sample standard deviations (SDs) of the paired differences. Panels b and c share the vertical range. Supplementary Fig. S4 shows additional transverse and longitudinal images.

## III. METHODS

### A. Specimens and immersion preparation

We analyzed two classes of unfilled glass vias fabricated by selective laser-induced chemical etching. The LR substrate has a nominal thickness of 150 µm and a via diameter of 50 µm, and the HR substrate has a nominal thickness of 500 µm and via diameter of 100 µm. Both were commissioned from SJ Company (Anyang-si, Gyeonggi-do, Republic of Korea) through Tomocube Inc. HR and LR denote these two tested substrate classes.

To improve RI reconstruction quality, the vias were infiltrated with commercial RI matching oils (Cargille Laboratories, Series A) and mounted between #1.5H coverslips for transmission imaging. Reducing RI contrast at the via wall brings the measurement closer to the weak-scattering condition assumed in reconstruction [40]. The immersion series spanned oil indices 1.490–1.510 for LR and 1.460–1.480 for HR, together with air and toluene (Supplementary Fig. S1). The selected oil indices were 1.500 for LR and 1.465 for HR.

### B. Intensity acquisition and refractive index reconstruction

For HT, we used a commercial HT-T1 Desktop (HT-T1D; Tomocube) system with 450 nm LED illumination, an illumination-pattern numerical aperture (NA) of 0.68 and a collection NA of 0.7. Four patterned-illumination images were acquired at each axial scan position with a configured increment of 0.703 µm (Fig. 1b). Intensity stacks were normalized by the corresponding background images and reconstructed by 3D PSF deconvolution with cosine pupil apodization and a Tikhonov coefficient of 10.0 [22], [23], [24]. Supplementary Methods specify the stored illumination patterns and transfer-function conventions.

TV refinement was applied to the dimensionless scattering potential $V = n^2/n_{bg}^2 - 1$, where $n$ is the RI and $n_{bg}$ is the calculation-background index. The graphics processing unit implementation used the fast iterative shrinkage-thresholding algorithm [41] with 100 outer iterations. Each iteration applied a gradient step constraining Fourier coefficients to those of the initial reconstruction within model-defined support, followed by a TV proximal operation using an $l_2$ discrete-gradient norm and 100 inner iterations. The TV weight was $\lambda_{\mathrm{TV}}=10^{-3}$ and the gradient-step parameter was 1.0. These settings implement TV-based refinement [42], [43] with constraints used in limited-angle tomography [36], [38]. The discrete-gradient and boundary conventions are specified in Supplementary Methods. Supplementary Fig. S2 compares TV weights; the selected settings were retained for the specimen comparisons.

The calculation-background indices were 1.470 for HR and 1.507 for LR at the 0.45 µm reconstruction wavelength. The axial grid interval was defined as the configured stage increment multiplied by these indices, giving stored axial grid intervals of 1.03341 and 1.059421 µm, respectively. Initial lateral sampling was 0.09 µm. Fourier-space cropping toward wavelength/(4NA) gave representative final lateral intervals of 0.160683 µm for HR and 0.160557 µm for LR; statistical processing used the rounded value 0.16 µm. These intervals specify the reconstruction and analysis grids.

### C. Boundary extraction and synthetic benchmark

The via boundary was extracted from two representations of the same reconstruction. The RI volume defines the lateral wall, because the glass and the infiltrated liquid differ in RI across the wall. At the axial openings, however, the liquid continues beyond the glass surfaces, so RI segmentation alone extends into the surrounding immersion medium (Supplementary Fig. S3). We used MP-HT to highlight the via boundary [41].

The lateral boundary was segmented using slice-wise Otsu thresholds on the RI volume [44], smoothed along depth with a Gaussian SD of two slices. Slices with thresholds above the calculation-background index were cleared. Border-component removal, hole filling and largest-component selection followed. MP-HT segmentation defined the retained axial span using the region of interest array's central lateral line; Supplementary Methods give the endpoint rule.

Canny edge extraction [45] used a Gaussian sigma of 1.4 pixels and thresholds of 0.1 and 0.2 on the binary cross-sections. After hourglass fitting, the optical surface was resampled at 360 angular positions per cross-section, with local-neighbor filling of empty angular bins. Each cross-section of this surface is termed a ring.

We evaluated boundary extraction using a synthetic cylindrical via 50 µm in diameter and 250 µm in axial length, with via-interior and background RI values of 1.465 and 1.470. The simulated reconstruction was analyzed using RI segmentation and MP-HT-assisted endpoint localization (Fig. 2c). Supplementary Methods specify the textured geometry and forward model. IoU was calculated over the full 3D segmented mask and reference mask:

$$IoU=\frac{|B_{\mathrm{seg}}\cap B_{\mathrm{ref}}|}{|B_{\mathrm{seg}}\cup B_{\mathrm{ref}}|}.$$

### D. Centerline and hourglass geometry

Circles were fitted to the segmented boundaries at each retained axial plane. Ordinary least-squares regression of their centers defined a straight centerline:

$$x_{\mathrm{c}}(z)=a_x+b_x z,\; y_{\mathrm{c}}(z)=a_y+b_y z.$$

The two functions give the fitted centerline coordinates in the original transverse scan planes at axial coordinate z.

The axial coordinate used for fitting the diameter profile was $\tilde{z}=z\sqrt{1+b_x^2+b_y^2}$, which accounts for centerline tilt. The boundary coordinates themselves remained in the original scan planes. Local circle diameters were fitted to a piecewise-linear hourglass profile with candidate waist coordinate $\tilde{z}_0$:

$$D_{\mathrm{fit}}(\tilde{z};\tilde{z}_0)=\begin{cases} D_0+m_{\mathrm{bottom}}(\tilde{z}-\tilde{z}_0), & \tilde{z}\le\tilde{z}_0, \\ D_0+m_{\mathrm{top}}(\tilde{z}-\tilde{z}_0), & \tilde{z}>\tilde{z}_0. \end{cases}$$

Candidate waist positions excluded the first and last five slices. At each candidate position, ordinary least squares estimated the waist diameter and the two branch slopes. Positive bottom slopes and negative top slopes were set to zero after fitting; the regression was not repeated after this step. The waist position was selected by the lowest resulting mean squared diameter residual. Half of the fitted diameter profile defined the reference radius.

Top and bottom diameters were the local circle diameters at the segmented endpoints. The reported waist diameter was the local circle diameter at the selected waist slice, rather than the fitted waist parameter $D_0$. Axial length was the difference between the endpoint scan coordinates. The upper and lower diameter slopes, $m_{\mathrm{top}}$ and $m_{\mathrm{bottom}}$, correspond to radius slopes $m/2$. Taper angles were calculated as

$$\theta_{\mathrm{top}}=90^\circ-\arctan(m_{\mathrm{top}}/2),$$
$$\theta_{\mathrm{bottom}}=90^\circ+\arctan(m_{\mathrm{bottom}}/2),$$

with arctangents in degrees. The factor of two converts diameter slope to radius slope. Angles are relative to the transverse plane, so a wall parallel to the via axis approaches 90°.

### E. Residual height, L-filter and sidewall roughness

The measured radius at each depth was the lateral-plane distance from a boundary point to the fitted centerline. Subtraction of the hourglass reference radius gave the residual height:

$$h(\theta_i, z_j) = r_{\text{measured}}(\theta_i, z_j) - r_{\text{fit}}(z_j).$$

In terms of the scan-plane coordinates of a surface point,

$$h_{ij} = \sqrt{[x_{ij} - x_{\text{c}}(z_j)]^2 + [y_{ij} - y_{\text{c}}(z_j)]^2} - r_{\text{fit}}(z_j).$$

A Gaussian L-filter following ISO 16610-61:2015, which specifies Gaussian filtration of nominal cylindrical surfaces, separated gradual radius changes, center offsets and ovality remaining in $h$ from finer wall texture. The waviness surface $w$ was obtained by low-pass filtering $h$ around the circumference and along the axis using the transmission

$$W(k) = 2^{-(k/k_{\text{c}})^2}, \quad k_{\text{c}} = 2\pi/\lambda_{\text{c}},$$

which transmits 50% of the amplitude at cut-off wavelength $\lambda_{\text{c}}$. Circumferential filtering used the periodic angular spectrum of each ring, computed by fast Fourier transform. The angular harmonic order $n$ was mapped to arc-length wavenumber $n/r_{\text{fit}}(z_j)$, giving $W_n(z_j) = 2^{-[n/(k_{\text{c}} r_{\text{fit}}(z_j))]^2}$. Axial filtering used a Gaussian kernel with SD $\sigma_z = \sqrt{2\ln 2}/k_{\text{c}} \approx 0.187\,\lambda_{\text{c}}$. Near the endpoints, normalized convolution renormalized the kernel over the available rings. The filtered height was

$$h_{\text{r}}(\theta_i, z_j) = h(\theta_i, z_j) - w(\theta_i, z_j).$$

The cut-off was defined by $\lambda_{\text{c}} = 2\pi R_{\text{nom}}/n_{\text{c}}$, with $n_{\text{c}} = 4$ undulations per revolution and nominal radius $R_{\text{nom}}$ of 50 µm for HR and 25 µm for LR. This gave $\lambda_{\text{c}}$ = 78.5 and 39.3 µm, respectively. Low-order harmonics $n \le 2$ describe slow radius changes, center offsets and ovality. At $r_{\text{fit}} = R_{\text{nom}}$, the waviness filter transmits approximately 96% and 84% of the $n=1$ and $n=2$ amplitudes and 50% at $n=4$. This low cut-off assigns local form to waviness while retaining higher-order sidewall texture. The physical cut-off $\lambda_{\text{c}}$ was fixed within each class, so its angular-harmonic equivalent varied with local reference radius. Sidewall roughness was the mean absolute value of the filtered height $h_{\text{r}}$ over all rings and angular samples:

$$S_{\text{a}} = \frac{1}{N_z N_\theta} \sum_{j=1}^{N_z} \sum_{i=1}^{N_\theta} |h_{\text{r}}(\theta_i, z_j)|.$$

For optical surfaces, each retained ring had equal weight, with 360 angular samples per ring. Rings within $\lambda_{\text{c}}/2$ of either endpoint were included using the normalized axial convolution described above. The reported $S_{\text{a}}$ applies the arithmetic mean height concept of ISO 25178-2:2021 to radial height deviations of the internal wall. Averaging uses a uniform angular–axial grid and the stated class-specific L-filter, without an additional short-wavelength (S) filter. This specifies the evaluation geometry and spatial band for comparison with other surface-texture measurements [12], [46].

### F. Paired comparisons and micro-computed tomography

To compare orientations, we measured the same specimens in the reference and physically flipped configurations. The reference cohort contained 100 HR and 92 LR vias; the flipped cohort contained 99 HR and 93 LR vias. Matching by via identifier gave 99 HR and 89 LR inversion pairs. The micro-CT cohort included 100 HR and 36 LR vias, of which 100 HR and 34 LR matched the reference cohort.

Micro-CT measurements used a commercial Baker Hughes v|tome|x s system with datos|x acquisition software. The source settings were 80 kV and 110 µA, with 2,000 configured projections over 360° and a detector timing setting of approximately 500 ms. Reconstructed volumes had a voxel spacing of 2.25 µm in both classes.

Micro-CT volumes underwent Otsu segmentation, Canny extraction, circle and hourglass fitting. Surface determination is a recognized component of micro-CT dimensional metrology [47], [48], [49], [50]. Without MP-HT, axial mask changes exceeding 10% of the maximum defined candidate bounds; the nearest bounds on either side of the volume center determined the retained span. Final endpoints were the first and last occupied slices. The micro-CT surface was represented by 300 angular samples per HR ring and 120 per LR ring; these are surface-sampling counts, not X-ray projection counts. The same class-specific L-filter cut-off $\lambda_{\text{c}}$ was used for optical and micro-CT surfaces. The roughness average used the corresponding angular sample count for each modality and class.

Bland–Altman plots [51], [52] show the mean of the reference-HT and comparison roughness on the horizontal axis and their difference (reference-HT − comparison) on the vertical axis. Bias is the mean paired difference; the 95% LoA are bias ±1.96 times the sample SD of the paired differences. All matched pairs were retained, including those outside these limits. Numerical summaries use means and between-via sample SDs.

## IV. DISCUSSION

HT makes the internal surfaces of intact TGVs accessible for both geometric and wall-texture measurements. HR walls were more nearly axial, whereas LR vias had a more pronounced hourglass profile. These two sampled substrate classes also showed different roughness distributions, with higher mean roughness in HR.

Holographic TGV inspection addresses morphology and roughness [16], while shadowgraphy, optical coherence tomography and spectral-coherence methods provide internal dimensions [17], [18], [19]. Here, extracting the full via boundary from an RI volume provides a common surface representation for inspecting the openings, waist and sidewall.

HT-based inspection has been applied to lenses, indented glass and volume gratings [33], [34], [35]. The present workflow extends volumetric RI-based inspection to etched TGVs. MP-HT provides complementary axial information for boundary extraction [39]; its use in other industrial specimens remains to be evaluated.

The L-filter separates gradual departures from the hourglass reference from finer wall structure. The maps locate this structure and scalar sidewall roughness summarizes its mean absolute amplitude. This construction applies the arithmetic mean height concept of ISO 25178-2:2021 to the curved internal wall using the specified radial reference and the cylindrical Gaussian filtration of ISO 16610-61:2015. Reported associations of glass texture with copper adhesion [12] and sidewall roughness with microwave behavior [13], [14] motivate inspection before metallization.

Little pooled bias after inversion does not establish absolute accuracy or same-condition repeatability. HT–micro-CT differences changed sign between the two classes, so the pooled bias is not a universal inter-modality correction. Sampling intervals, synthetic volume overlap and bulk caliper comparisons do not independently calibrate surface-height accuracy. The tested vias also do not establish variability across independently replicated fabrication lots.

Volatile index-matching media offer a route to simplify post-imaging removal and reduce persistent oil residues. Toluene, included in our immersion comparison, is a candidate for this approach; mixing it with isopropyl alcohol could further tune the immersion-medium RI for different glass substrates. Such solvent-based formulations could facilitate integration of optical inspection with subsequent substrate processing.

The extracted surface provides geometric and texture descriptors before metallization while preserving the intact via. In the context of laser modification and selective etching [3], [4], [5], [6], [7], [8], [9], [10], [11], these descriptors characterize post-etch morphology. Complementary work has reported nondestructive characterization of laser-induced modification in glass before selective etching [51]. Inspection after metallization or bonding is a possible future direction, but opaque layers limit visible-light transmission and would require a different imaging configuration.

## V. CONCLUSION

We used HT to measure the full 3D geometry and internal wall texture of intact, unfilled TGVs nondestructively. Complementary RI and MP-HT representations localized the internal boundary for measurements of local diameters, taper, axial length and sidewall texture. Measurements of 192 vias demonstrate the workflow across two substrate classes. Physical inversion showed little average shift in roughness, while micro-CT comparisons revealed class-dependent modality differences. The method makes internal geometric and wall-texture descriptors available before copper filling.

## DATA AND CODE AVAILABILITY

Data and analysis code supporting the findings of this study are available from the corresponding author upon reasonable request.

## AUTHOR CONTRIBUTIONS AND DECLARATIONS

Dohyeon Lee performed the measurements and experiments and analyzed the data. Herve Hugonnet contributed to data analysis and provided analysis tools. YongKeun Park conceived the study and acquired funding. All authors contributed to writing the manuscript. YongKeun Park is affiliated with Tomocube, which supplied or commissioned the specimens used in this study.

## ACKNOWLEDGMENT

This work was supported by the Samsung Research Funding Center of Samsung Electronics under Grant SRFC-IT2501-05 and by the National Research Foundation of Korea grant funded by the Korea government (MSIT) (RS-2022-NR068141). YongKeun Park acknowledges support from the Merck fellowship.

OpenAI Codex assisted manuscript editing and proofreading. The authors reviewed the content and take responsibility for the final manuscript.

DOHYEON LEE is affiliated with the Department of Physics and the Institute of Imaging, Korea Advanced Institute of Science and Technology, Daejeon, Republic of Korea.

HERVE HUGONNET is affiliated with the Department of Physics and the Institute of Imaging, Korea Advanced Institute of Science and Technology, Daejeon, Republic of Korea.

YONGKEUN PARK is affiliated with the Department of Physics and the Institute of Imaging, Korea Advanced Institute of Science and Technology, and Tomocube, Daejeon, Republic of Korea.

**Supplementary information for through-glass via holotomography**

Dohyeon Lee, Herve Hugonnet and YongKeun Park

**Supplementary numerical summaries**

**Supplementary Table S1 Population roughness and axial length**

| Quantity | High-roughness (HR) (*N* = 100) | Low-roughness (LR) (*N* = 92) |
|---|---|---|
| Sidewall roughness (µm), mean ± sample standard deviation (SD) | 0.841 ± 0.028 | 0.272 ± 0.098 |
| Segmented axial length (µm), mean ± sample SD | 516.8 ± 9.7 | 157.1 ± 5.8 |
| Separate bulk caliper reading (µm) | 520 | 140 |
| Optical mean − caliper reading (µm) | −3.2 | +17.1 |
| Difference relative to bulk reading (%) | −0.6 | +12.2 |

*N* denotes the number of vias measured in the reference orientation. The SD describes between-via variation. Bulk caliper readings were not co-located with individual vias; no independently established caliper uncertainty is assigned. These comparisons do not establish traceable axial accuracy.

**Supplementary Table S2 Paired roughness comparisons**

| Quantity | Reference holotomography (HT) versus flipped HT | Reference HT versus micro-computed tomography (micro-CT) |
|---|---|---|
| Matched pairs (high-roughness (HR); low-roughness (LR)) | 188 (HR = 99; LR = 89) | 134 (HR = 100; LR = 34) |
| Pooled mean difference (µm) | +0.002 | +0.051 |
| 95% limits of agreement (LoA) (µm) | −0.094 to +0.097 | −0.223 to +0.325 |
| Pairs outside pooled limits | 6 | 11 |
| HR mean difference (µm) | +0.008 | +0.124 |
| LR mean difference (µm) | −0.005 | −0.164 |
| HR paired means: reference; comparison (µm) | 0.841; 0.833 | 0.841; 0.717 |
| LR paired means: reference; comparison (µm) | 0.272; 0.277 | 0.272; 0.436 |

Mean differences are reference HT minus the comparison measurement. Paired means are listed in the order reference HT; comparison (flipped HT or micro-CT, according to the column). The 95% LoA are the pooled bias ±1.96 times the sample standard deviation (SD) of the paired differences. All matched pairs are retained. The LR micro-CT cohort contained 36 measured vias, of which 34 matched the HT reference population. Reported means and differences are independently rounded, so their last displayed digits need not subtract exactly. Paired means use only the matched cohort; population means in Table S1 use the full reference cohort.

**Supplementary Methods**

**Illumination and reconstruction.** The four illumination distributions were stored as a 1200 × 1200 × 4 array. Each distribution was resampled onto the reconstruction Fourier grid, restricted to the illumination aperture, clipped to nonnegative values and normalized to unit mean. The patterned-source calculation used the wavelength, numerical aperture (NA) values and cosine pupil apodization stated in Methods B to construct phase and amplitude transfer components for point-spread function (PSF) deconvolution. Fourier support was defined by the aperture geometry, calculation-background refractive index (RI), wavelength and grid. The stored arrays and implementation are covered by the main-text Data and Code Availability statement.

**Total variation (TV) refinement.** The TV weight was $\lambda_{\mathrm{TV}} = 10^{-3}$ and the gradient-step parameter was 1.0. These settings implement TV-based refinement [1], [2] with constraints used in limited-angle tomography [3], [4]. Before each TV proximal step, the Fourier coefficients of the current scattering-potential estimate inside the model-defined support were set to those of the initial reconstruction; coefficients outside this support were unchanged by that constraint. This implements the support-gradient update with a step parameter of 1.0. The TV step used an isotropic $l_2$-norm TV proximal operator ($p$ = 2) and the main-text iteration counts and weight, followed by fast iterative shrinkage-thresholding acceleration. The discrete differences used adjacent array indices with equal directional weighting and periodic array boundaries. They were not divided by the lateral or axial voxel intervals; the weight therefore applies to this dimensionless grid convention.

The calculation-background RI values were 1.470 for high-roughness (HR) and 1.507 for low-roughness (LR) vias at the 0.45 µm reconstruction wavelength. The axial grid interval was defined as the configured stage increment multiplied by these indices, giving axial grid intervals of approximately 1.03 and 1.06 µm, respectively. Initial lateral sampling was 0.09 µm. Fourier-space cropping toward wavelength/(4NA) gave representative final lateral intervals of approximately 0.16 µm for both HR and LR; statistical processing used the rounded value 0.16 µm. These intervals specify the reconstruction and analysis grids.

**Axial coordinates.** The axial grid interval was defined as the configured stage increment multiplied by the calculation-background RI, 1.470 for HR and 1.507 for LR, giving saved axial intervals of 1.03341 µm for HR and 1.059421 µm for LR. The analysis retained these saved coordinates for axial lengths, taper fits and axial filtering. The calculation-background RI is distinct from the RI of the selected immersion liquid. This coordinate convention does not constitute an independent axial calibration.

**Optical endpoints.** For holotomography (HT), the morphology-preserving holotomography (MP-HT) calculation used three reflected slices at each axial boundary. Lateral Gaussian smoothing used $\sigma_{xy} = \frac{1}{8 b \Delta x}$ pixels, where $b$ is the axial half-width of the morphology-transfer support in cycles per micrometer and $\Delta x$ is the lateral grid interval in micrometers. A global Otsu threshold of the final MP-HT volume defined an edge mask. Components touching a lateral border were removed. Holes were filled in each slice, and the original edge pixels were then removed to isolate the enclosed interior. The first and last voxels along the central lateral line of the region-of-interest array defined the axial crop of the RI mask. This line was the array midpoint, not the fitted centerline. The first and last slices with nonzero mask area then defined the final endpoints.

**Micro-computed tomography endpoints.** For micro-computed tomography (micro-CT), axial endpoints were determined from changes in the segmented mask without MP-HT endpoint localization. The algorithm summed the absolute difference between consecutive binary-mask slices over both lateral coordinates. Changes greater than 0.1 times the maximum were flagged. The last flagged change before the volume midplane and the first after it bounded the central span; slices outside this span were cleared. Final endpoints were the first and last retained slices with nonzero mask area. These algorithmic rules define the measured axial extent; the bulk comparison in Fig. S5 is separate.

**Boundary extraction.** Canny edge extraction [5] used a Gaussian standard deviation (SD) of 1.4 pixels and thresholds of 0.1 and 0.2 on the binary cross-sections. After hourglass fitting, the optical surface was resampled at 360 angular positions per cross-section, with local-neighbor filling of empty angular bins. Each cross-section of this surface is termed a ring.

We evaluated boundary extraction using a synthetic cylindrical via 50 µm in diameter and 250 µm in axial length, with via-interior and background RI values of 1.465 and 1.470. The simulated reconstruction was analyzed using RI segmentation and MP-HT-assisted endpoint localization (Fig. 2c). Intersection over union (IoU) was calculated from the full three-dimensional (3D) segmented mask and reference mask:

$$IoU = \frac{|B_{\text{seg}} \cap B_{\text{ref}}|}{|B_{\text{seg}} \cup B_{\text{ref}}|}.$$

**Centerline and hourglass geometry.** Circles were fitted to the segmented boundaries at each retained axial plane. Ordinary least-squares regression of their centers defined a straight centerline:

$$x_{\text{c}}(z) = a_x + b_x z,\ y_{\text{c}}(z) = a_y + b_y z.$$

The two functions give the fitted centerline coordinates in the original transverse scan planes at axial coordinate z.

The axial coordinate used for fitting the diameter profile was $\tilde{z} = z\sqrt{1 + b_x^2 + b_y^2}$, which accounts for centerline tilt. The boundary coordinates themselves remained in the original scan planes. Local circle diameters were fitted to a piecewise-linear hourglass profile with candidate waist coordinate $\tilde{z}_0$:

$$D_{\text{fit}}(\tilde{z};\tilde{z}_0) = \begin{cases} D_0 + m_{\text{bottom}}(\tilde{z} - \tilde{z}_0), & \tilde{z} \le \tilde{z}_0, \\ D_0 + m_{\text{top}}(\tilde{z} - \tilde{z}_0), & \tilde{z} > \tilde{z}_0. \end{cases}$$

Candidate waist positions excluded the first and last five slices. At each candidate position, ordinary least squares estimated the waist diameter and the two branch slopes. Positive bottom slopes and negative top slopes were set to zero after fitting; the regression was not repeated after this step. The waist position was selected by the lowest resulting mean squared diameter residual. Half of the fitted diameter profile defined the reference radius.

Top and bottom diameters were the local circle diameters at the segmented endpoints. The reported waist diameter was the local circle diameter at the selected waist slice, rather than the fitted waist parameter $D_0$. Axial length was the difference between the endpoint scan coordinates. The upper and lower diameter slopes, $m_{\text{top}}$ and $m_{\text{bottom}}$, correspond to radius slopes $m/2$. Taper angles were calculated as

$$\theta_{\text{top}} = 90^\circ - \arctan(m_{\text{top}}/2),$$

$$\theta_{\text{bottom}} = 90^\circ + \arctan(m_{\text{bottom}}/2),$$

with arctangents in degrees. The factor of two converts diameter slope to radius slope. Angles are relative to the transverse plane, so a wall parallel to the via axis approaches 90°.

**Residual height, L-filter and sidewall roughness**. The measured radius at each depth was the lateral-plane distance from a boundary point to the fitted centerline. Subtraction of the hourglass reference radius gave the residual height:

$$h(\theta_i, z_j) = r_{\text{measured}}(\theta_i, z_j) - r_{\text{fit}}(z_j).$$

In terms of the scan-plane coordinates of a surface point,

$$h_{ij} = \sqrt{[x_{ij} - x_{\text{c}}(z_j)]^2 + [y_{ij} - y_{\text{c}}(z_j)]^2} - r_{\text{fit}}(z_j).$$

A Gaussian L-filter following ISO 16610-61:2015, which specifies Gaussian filtration of nominal cylindrical surfaces, separated gradual radius changes, center offsets and ovality remaining in $h$ from finer wall texture. The waviness surface $w$ was obtained by low-pass filtering $h$ around the circumference and along the axis using the transmission

$$W(k) = 2^{-(k/k_{\text{c}})^2}, \quad k_{\text{c}} = 2\pi/\lambda_{\text{c}},$$

which transmits 50% of the amplitude at cut-off wavelength $\lambda_{\text{c}}$. Circumferential filtering used the periodic angular spectrum of each ring, computed by fast Fourier transform. The angular harmonic order $n$ was mapped to arc-length wavenumber $n/r_{\text{fit}}(z_j)$, giving $W_n(z_j) = 2^{-[n/(k_{\text{c}} r_{\text{fit}}(z_j))]^2}$. Axial filtering used a Gaussian kernel with SD $\sigma_z = \sqrt{2\ln 2}/k_{\text{c}} \approx 0.187\,\lambda_{\text{c}}$. Near the endpoints, normalized convolution renormalized the kernel over the available rings. The filtered height was

$$h_r(\theta_i, z_j) = h(\theta_i, z_j) - w(\theta_i, z_j).$$

The cut-off was defined by $\lambda_c = 2\pi R_{nom}/n_c$, with $n_c = 4$ undulations per revolution and nominal radius $R_{nom}$ of 50 µm for HR and 25 µm for LR. This gave $\lambda_c$ = 78.5 and 39.3 µm, respectively. Low-order harmonics $n \le 2$ describe slow radius changes, center offsets and ovality. At $r_{fit} = R_{nom}$, the waviness filter transmits approximately 96% and 84% of the $n = 1$ and $n = 2$ amplitudes and 50% at $n = 4$. This low cut-off assigns local form to waviness while retaining higher-order sidewall texture. The physical cut-off $\lambda_c$ was fixed within each class, so its angular-harmonic equivalent varied with local reference radius. Sidewall roughness was the mean absolute value of the filtered height $h_r$ over all rings and angular samples:

$$S_a = \frac{1}{N_z N_\theta} \sum_{j=1}^{N_z} \sum_{i=1}^{N_\theta} |h_r(\theta_i, z_j)|.$$

For optical surfaces, each retained ring had equal weight, with 360 angular samples per ring. Rings within $\lambda_c/2$ of either endpoint were included using the normalized axial convolution described above. The reported $S_a$ applies the arithmetic mean height concept of ISO 25178-2:2021 to radial height deviations of the internal wall. Averaging uses a uniform angular–axial grid and the stated class-specific L-filter, without an additional short-wavelength (S) filter. This specifies the evaluation geometry and spatial band for comparison with other surface-texture measurements [6], [7].

**Surface extraction and resampling.** Slice-wise RI Otsu thresholds were smoothed along depth with a Gaussian SD of two slices. Canny edges were fitted and resampled at 360 angular positions per optical ring, with local-neighbor filling of empty bins. Optical and micro-CT sampling and the roughness definition are specified in Methods C–F of the main text.

**Synthetic geometry.** The benchmark used a synthetic cylindrical via 50 µm in diameter and 250 µm in axial length, with via-interior and background RI values of 1.465 and 1.470. Texture was introduced by 500 spherical extensions of the via interior, each 15 µm in diameter, placed at uniformly sampled angular and axial positions on the cylindrical wall using random seed 42. Extensions were restricted to the axial slab. The grid contained 498 × 498 × 339 voxels, with lateral and axial spacings of 0.1606626506 and 1.03341 µm. Otsu segmentation and component cleanup of the ideal RI volume supplied the reference mask.

**Synthetic imaging.** The forward simulation used a wavelength of 0.45 µm, a collection NA of 0.7, an illumination-pattern NA of 0.68, cosine pupil apodization and no added signal noise. Inverse-PSF reconstruction used a Tikhonov coefficient of 10.0. A pure-phase multi-pattern forward model generated the intensity data for reconstruction and TV refinement. The resulting IoU values were 0.93 for RI segmentation and 0.96 with MP-HT-assisted endpoint localization. The 15 µm extensions define the test geometry, not an estimate of optical resolution.

**Micro-computed tomography preprocessing.** Reconstructed 16-bit micro-CT volumes were converted to Zarr, aligned to the via direction using first-order interpolation, and cropped into individual vias. The aligned volumes were then segmented and analyzed as described in Methods F of the main text.

**Micro-computed tomography comparison.** Micro-CT volumes underwent Otsu segmentation, Canny extraction, circle and hourglass fitting. Surface determination is a recognized component of micro-CT dimensional metrology [8], [9], [10], [11]. Without MP-HT, axial mask changes exceeding 10% of the maximum defined candidate bounds; the nearest bounds on either side of the volume center determined the retained span. Final endpoints were the first and last occupied slices. The micro-CT surface was represented by 300 angular samples per HR ring and 120 per LR ring; these are surface-sampling counts, not X-ray projection counts. The same class-specific L-filter cut-off $\lambda_c$ was used for optical and micro-CT surfaces. The roughness average used the corresponding angular sample count for each modality and class.

Bland–Altman plots [12], [13] show the mean of the reference-HT and comparison roughness on the horizontal axis and their difference (reference-HT – comparison) on the vertical axis. Bias is the mean paired difference; the 95% limits of agreement (LoA) are bias ±1.96 times the sample SD of the paired differences. All matched pairs were retained, including those outside these limits. Numerical summaries use means and between-via sample SDs.

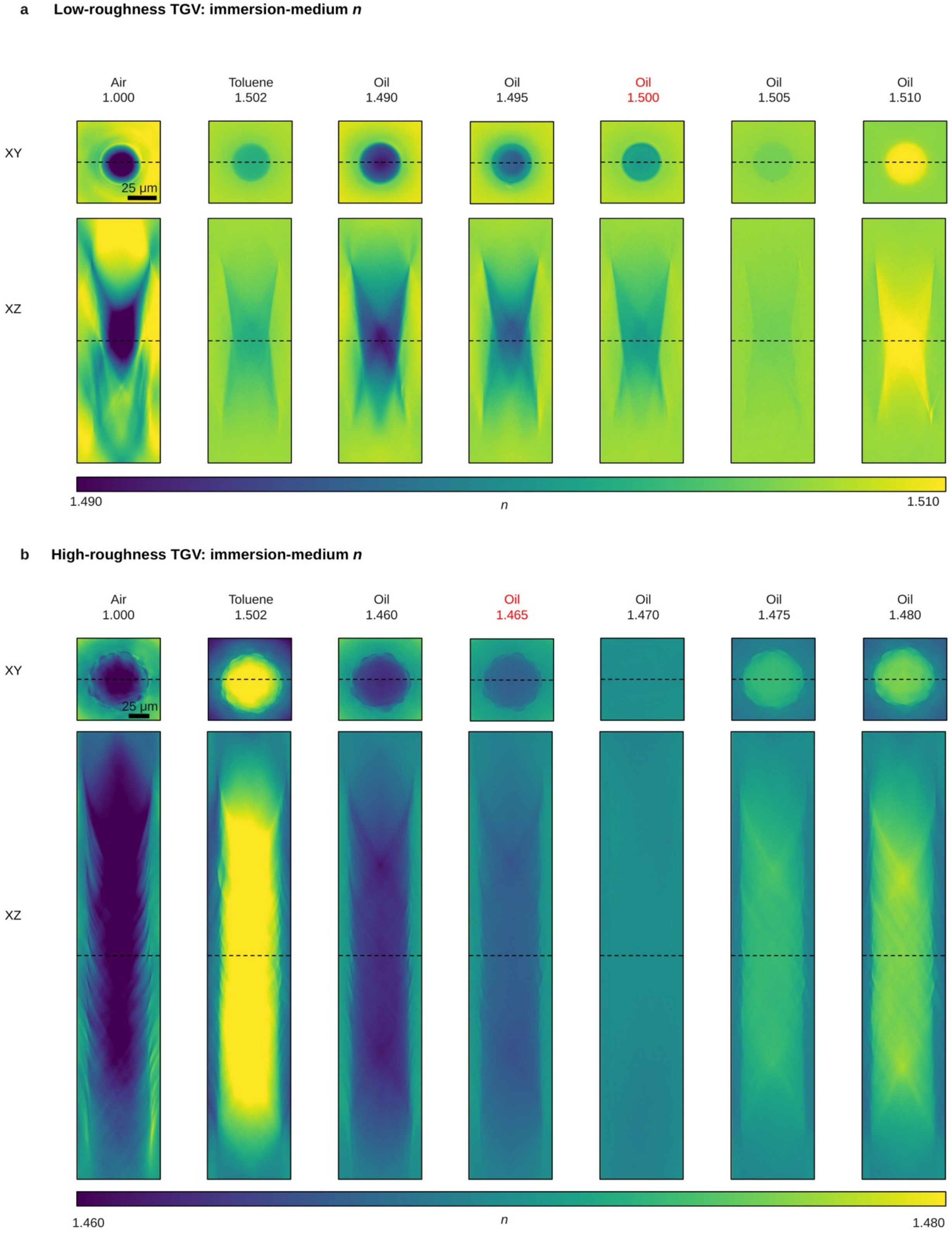

**Supplementary Figure S1.** Immersion conditions and boundary contrast in through-glass vias (TGVs). (a) Low-roughness (LR) examples in air, toluene ($n$ = 1.502) and index-matching oils with $n$ = 1.490, 1.495, 1.500, 1.505 and 1.510. (b) High-roughness (HR) examples in air, toluene ($n$ = 1.502) and index-matching oils with $n$ = 1.460, 1.465, 1.470, 1.475 and 1.480. Dashed lines mark the orthogonal sections. Red labels mark the selected oil conditions: 1.500 for LR and 1.465 for HR. Images acquired in air should not be interpreted as calibrated measurements because the large refractive index (RI) mismatch violates the weak-scattering assumption required for RI reconstruction. Scale bars, 25 µm. The symbol $n$ denotes the RI of the immersion medium; XY and XZ denote the transverse and longitudinal planes, respectively.

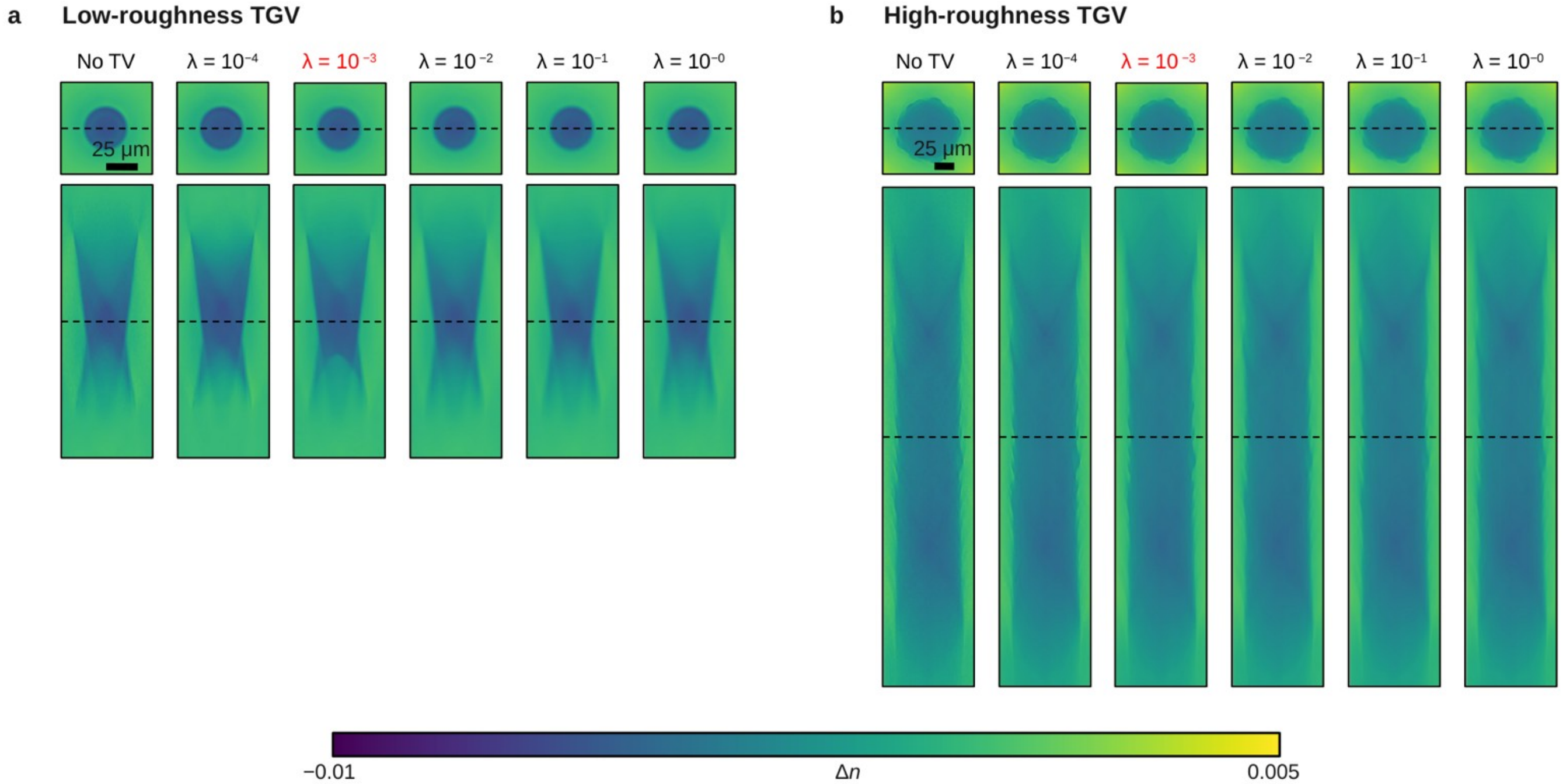

**Supplementary Figure S2.** Reconstruction dependence on total variation (TV) refinement. Low-roughness (LR) and high-roughness (HR) through-glass vias (TGVs) show the initial reconstruction without TV refinement and reconstructions with coefficients $10^{-4}$, $10^{-3}$, $10^{-2}$, $10^{-1}$ and $1$. Dashed lines mark the orthogonal sections. The selected coefficient, marked in red, is $\lambda_{TV} = 10^{-3}$. Scale bars, 25 µm. The refractive index (RI) contrast is $\Delta n = n - n_{bg}$, where $n$ is the RI and $n_{bg}$ is the reconstruction-background index (1.470 for HR and 1.507 for LR). In the panels, λ denotes the TV weight; XY and XZ denote the transverse and longitudinal planes, respectively.

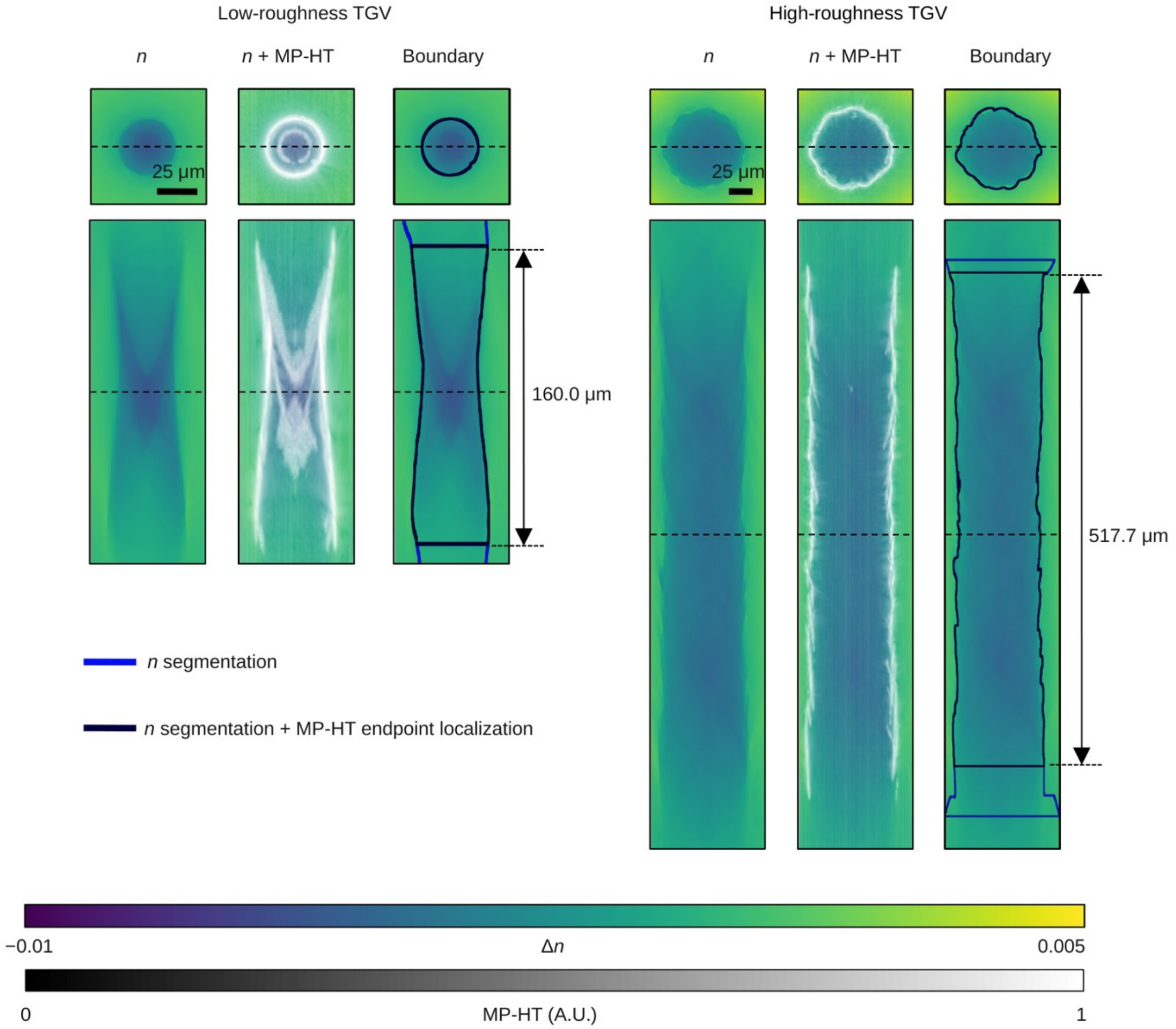

**Supplementary Figure S3.** Boundary extraction from experimental holotomography (HT) volumes for low-roughness (LR) and high-roughness (HR) through-glass vias (TGVs). For each class, transverse and longitudinal sections show the refractive index (RI) contrast, RI combined with the morphology-preserving holotomography (MP-HT) representation, and the extracted boundaries. RI segmentation alone (blue) extends into the surrounding immersion liquid at the axial ends, whereas MP-HT (dark blue) localizes the axial endpoints. Dashed lines mark the orthogonal sections. The segmented axial lengths are 160.0 µm for LR and 517.7 µm for HR. Scale bars, 25 µm. XY and XZ denote the transverse and longitudinal planes, respectively. The displayed contrast is $\Delta n = n - n_{bg}$, where $n$ denotes RI and $n_{bg}$ is the reconstruction-background index (1.470 for HR and 1.507 for LR). MP-HT intensity is shown in arbitrary units (A.U.).

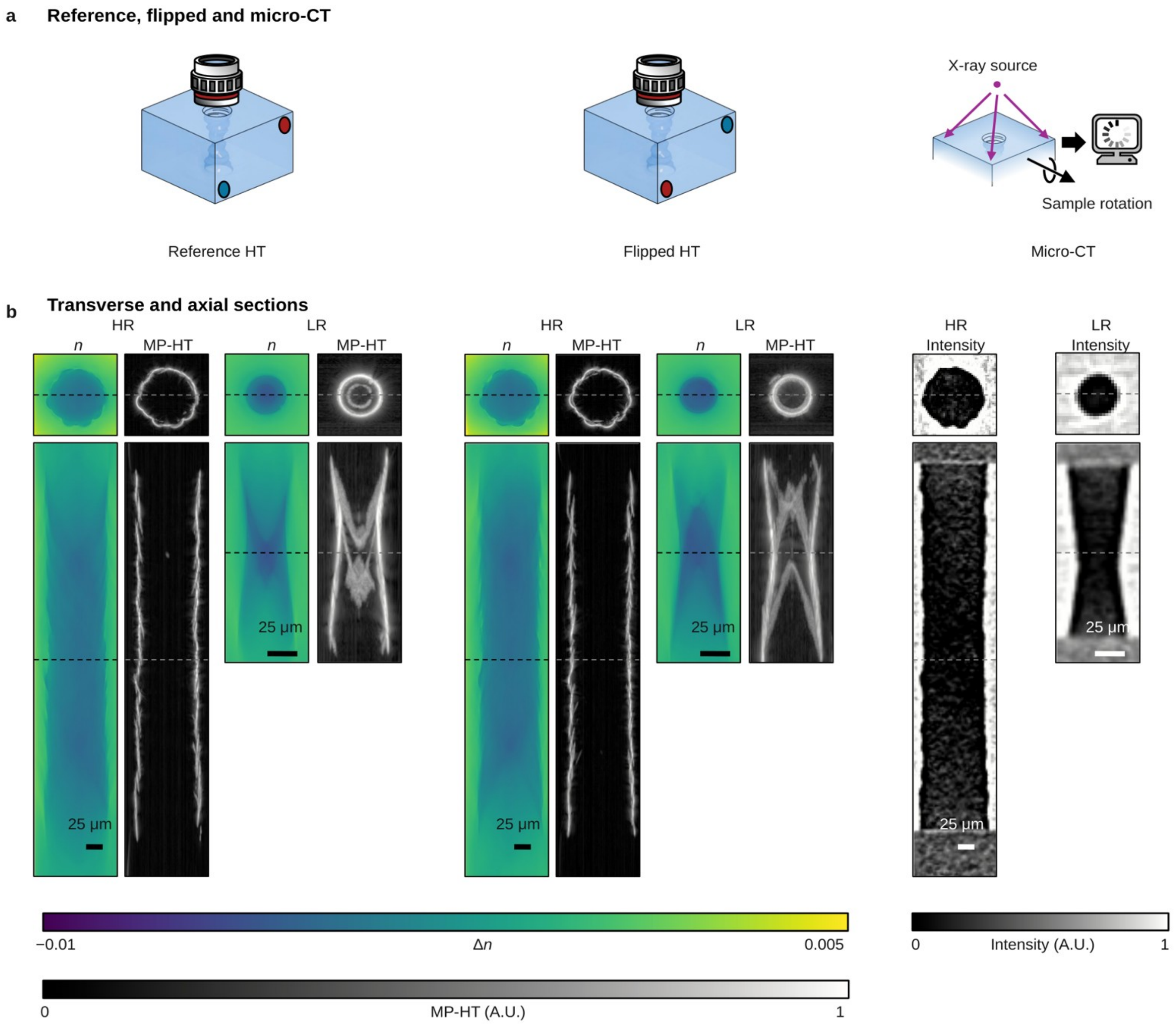

**Supplementary Figure S4.** Orientation and modality comparisons in transverse and longitudinal sections. Reference and physically flipped holotomography (HT) measurements are compared with micro-computed tomography (micro-CT) in the high-roughness (HR) and low-roughness (LR) classes. Refractive index (RI), morphology-preserving holotomography (MP-HT) and micro-CT views are shown with their corresponding color scales or grayscales. Dashed lines mark the orthogonal sections. Scale bars, 25 µm. The displayed contrast is $\Delta n = n - n_{bg}$, where $n$ denotes RI and $n_{bg}$ is the reconstruction-background index (1.470 for HR and 1.507 for LR). A.U. denotes arbitrary units.

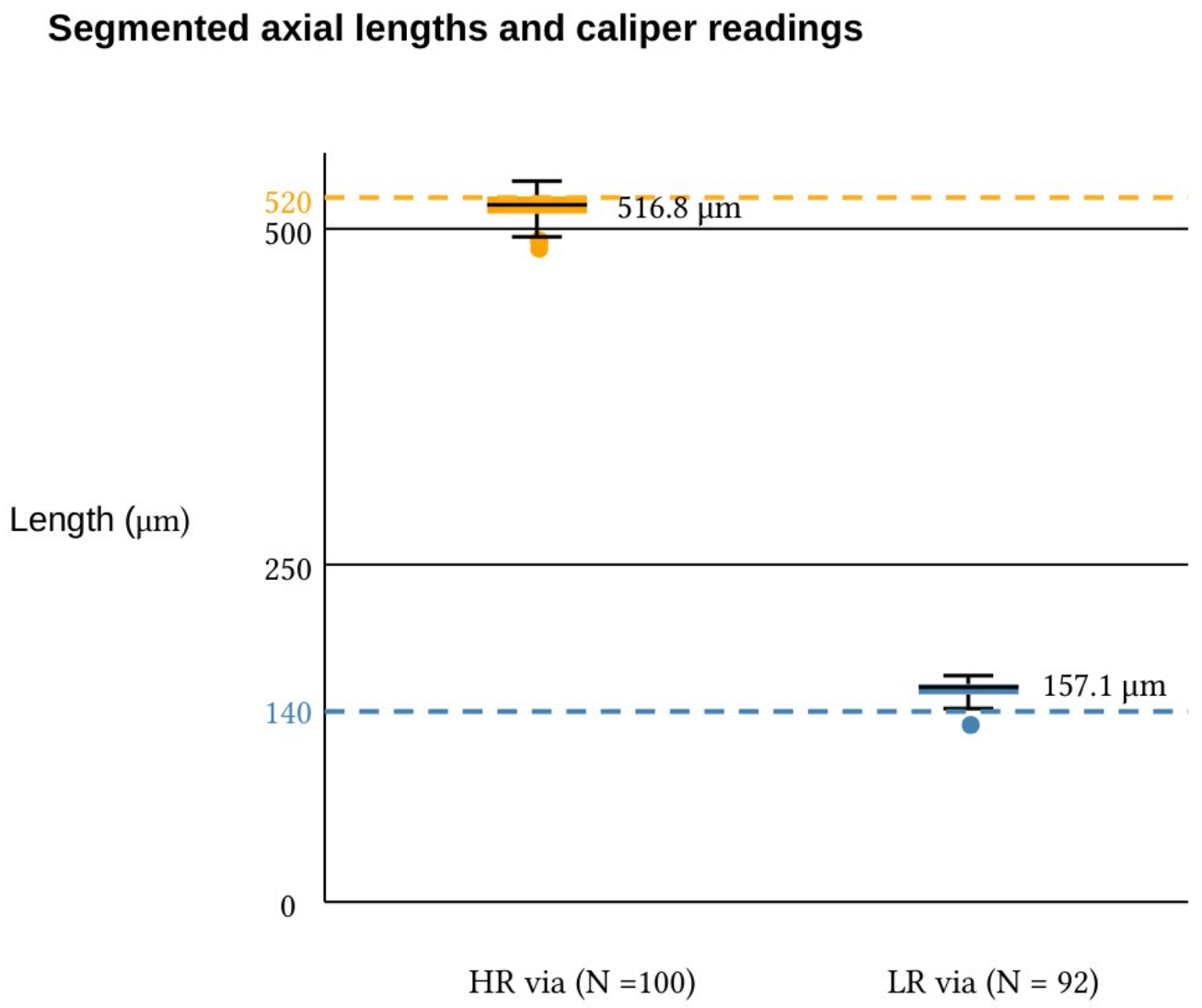

**Supplementary Figure S5.** Comparison of segmented axial lengths with bulk caliper readings. The mean and sample standard deviation (SD) of optical axial lengths are 516.8 ± 9.7 μm for high-roughness (HR) vias and 157.1 ± 5.8 μm for low-roughness (LR) vias. Caliper readings are 520 and 140 μm, respectively (dashed lines). Values and bulk-reading differences are listed in Supplementary Table S1. This comparison does not assign a co-located thickness or traceable uncertainty to an individual via. The symbol $N$ denotes the number of vias.